\documentstyle[preprint,pra,aps]{revtex}

\begin{document}

\bibliographystyle{prsty}

\draft

\preprint{to be published in Phys.~Rev.~A}

\tighten

\title{Quantum Noise and Polarization Fluctuations in \\
Vertical Cavity Surface Emitting Lasers}

\author{Holger F. Hofmann and Ortwin Hess}
\address{Institute of Technical Physics, DLR\\
Pfaffenwaldring 38--40, D--70569 Stuttgart, Germany}

\date{\today}

\maketitle

\begin{abstract}
We investigate the polarization fluctuations caused by quantum noise 
in quantum well vertical cavity surface emitting lasers (VCSELs). 
Langevin equations are derived 
on the basis of a generalized rate equation model in which the influence 
of competing gain-loss and frequency anisotropies is included. 
This reveals how the anisotropies and the quantum well confinement effects 
shape the correlations and the magnitude of fluctuations in ellipticity 
and in polarization direction.   
According to our results all parameters used in the rate equations 
may be obtained experimentally from precise time resolved measurements of the intensity and
polarization  fluctuations in the emitted laser light.
To clarify the effects of anisotropies and of quantum well confinement on the laser process  in
VCSELs we therefore propose time resolved measurements of the polarization fluctuations in the laser
light. 
In particular, such measurements allow to distinguish the effects of frequency anisotropy and of
gain-loss anisotropy and would provide data on the spin relaxation rate in the quantum well
structure during cw operation as well as representing a new way of experimentally determinig the
linewidth enhancement factor $\alpha$. 
\end{abstract}



\section{Introduction}

\label{sec:intro} 

Vertical Cavity Surface Emitting Lasers (VCSEL) 
have recently attracted  intensive experimental and theoretical effort.
One of the major advantages VCSELs have over conventional semiconductor 
lasers is the highly symmetric geometry around the axis of laser light 
emission. The polarization of light emitted from VCSELs is
therefore not determined by the massive anisotropy of the device
architecture, as is the case in edge 
emitting lasers. Instead, due to the transverse symmetry of the cavity and the
active region the polarization is highly sensitive to more
subtle effects, such as small anisotropies in the crystal structure, strain or
optical anisotropies in the mirrors 
\cite{Cha91,Koy91,Vak95,Cho95,Jan96}.

Experiments have already provided a large range of results on polarization 
as a function of pumping current \cite{Cho95,Cho94}. The variety of results 
indicates the 
sensitive dependence of polarization on very small effects. However, the 
connection between the observed stability or bistability of linear 
polarization and
the theoretical explanations offered are still no more than tentative. 

In 1995, San Miguel, Feng and Moloney introduced a rate equation model for
quantum well VCSELs, based on the observation, that the electron-hole pairs
in the bands closest to the band gap can be separated into electron-hole
pairs emitting only right circular polarized light and electron-hole pairs 
emitting only left circular polarized light
\cite{Mig95}. Effectively, this corresponds to 
two independent reservoirs of electron hole pairs, each characterized by its 
own electron-hole pair density, coupled only by spin relaxation processes. 
In the following, this assumption will be refered to as the split density
model.

The first major result derived from this model 
has been the proposal that polarization stability may be a consequence
of a frequency anisotropy due to birefringence in the optical cavity
\cite{Mar96}. As opposed to the more straightforward 
assumption of a gain-loss anisotropy, the frequency anisotropy may give 
rise to polarization switching as the injection current is increased.
However, since polarization switching may also be induced by temperature 
related effects \cite{Cho94}, 
the fact that polarization switching is experimentally observed is 
not sufficient to verify the assumption of dominating frequency anisotropies. 
Although further work based on the split density model of San Miguel and coworkers has
been recently forthcoming \cite{Jan96,Tra96}, 
the predictions
made by the model have so far been insufficient to truly rule out the 
presence of
alternative mechanisms. In particular, no attempts have yet been made to
determine the anisotropies and the time scales involved in the laser process
directly from the laser emission during cw operation.

In order to demonstrate that such a direct test of the model is indeed 
possible, we investigate the 
polarization fluctuations predicted by the Langevin equations derived from
the split density model. One of the advantages of this approach is 
that polarization fluctuations can be measured at a fixed injection current 
during cw operation. Deriving the anisotropies 
from fluctuation measurements provides not only a tool for testing
the theory, but can also provide information about temperature 
dependent changes in the anisotropies.
In this way, our theory may help to resolve questions such as those raised
in \cite{Cho94} and \cite{Tra96}. 

In section~\ref{sec:rate} of this paper, we formulate the full 
rate equations of the split density model, using the normalized  Stokes parameters to describe light
field polarization.  
In section~\ref{sec:Langevin}, we examine the most likely case of a frequency
anisotropy and a gain-loss anisotropy along the same or two orthogonal crystal axes. The linearized
Langevin equations for the stationary point are derived. 
In section~\ref{sec:solution}, the solution of the Langevin equation is given, using reasonable 
approximations with regard to the time scales involved. 
In section~\ref{sec:discussion}, the results are discussed and possible experiments are proposed. 
Finally, in section~\ref{sec:concl}, conclusions and an outlook are presented.    
     

\section{Formulation of the Rate Equations for the Split Density Model}

\label{sec:rate}

\subsection{The split density model}

If only the lowest lying conduction bands and the highest lying valence 
bands contribute
to the laser process in a quantum well structure, the preservation of
angular momentum around the axis perpendicular to the quantum well and 
the fact that photons which are emitted along this axis have 
angular momenta of $\pm 1$
(corresponding to either left or right circular polarization) limits the number
of possible emission processes to two separate transitions. Conduction
band electrons with a spin of +1/2 around the axis perpendicular to the well
can only recombine with heavy holes of -3/2 angular momentum around this axis,
emitting a photon with an angular momentum of -1. (Note, that a hole with an 
angular momentum of -3/2 corresponds to an empty electronic level with an
angular momentum of +3/2.) Correspondingly, spin -1/2 electrons only 
recombine with holes of +3/2 angular momentum, emitting photons with an 
angular momentum of +1. 

This means that effectively two completely distinct pools of
electron-hole pairs exist.
In one pool each electron-hole pair has an angular momentum of -1 
and interacts only with right circular polarized light.
The electron-hole pairs in the other pool, having an 
angular momentum of +1 per pair, interact only with left circular 
polarized light. 
Rate equations based on this assumption have 
originally been formulated by San Miguel and coworkers in
\cite{Mig95}. 
In the following we adopt the notion of the split density model and
considerably extend the analysis towards a generalized representation 
able to accomodate in the model arbitrary types of anisotropies.

\subsection{Parameters}
As in the paper by San Miguel \cite{Mig95}, we use the variables D for the total 
electron-hole density and d for the difference between the densities of 
electron-hole pairs with +1 and with -1 angular momentum. We will use a 
somewhat different approach to describe the light field in
the cavity, however, to emphasize the difference between light field intensity
and light field polarization. The total number of photons in the cavity
(regardless of polarization) is denoted by $n$. The polarization is then described by a three
dimensional vector of unit length, {\bf P}, which defines the point on the
Poincare sphere corresponding to the present polarization. The components of 
{\bf P} are the normalized Stokes parameters. 
If $E_\pm$ are the complex electric field amplitudes of right and
left circular light, the normalized Stokes parameters are given by
\begin{mathletters}
\begin{eqnarray}
\label{eq:stokespara}
P_1 & = & \frac{E_+^*E_- + E_-^*E_+}{E_+^*E_+ + E_-^*E_-}\\
P_2 & = & -i\frac{E_+^*E_- - E_-^*E_+}{E_+^*E_+ + E_-^*E_-}\\
P_3 & = & \frac{E_+^*E_+ - E_-^*E_-}{E_+^*E_+ + E_-^*E_-}
\end{eqnarray}
\end{mathletters}
The spatial directions of the vector {\bf P} correspond to pairs of 
orthogonal polarizations. The component of {\bf P} along any given direction
is equal to the intensity difference between the two polarizations 
relative to the total intensity. $P_3$ is the intensity difference between
left and right circular polarization, $P_1$ is the intensity difference 
between x and y polarization and $P_2$ is the intensity difference between
light polarized along the (1,1) direction and light polarized along
the (1,-1) direction. In the plane of linear polarization, a rotation of
{\bf P} by 180 degrees therefore corresponds to a 90 degree change in the 
direction of polarization. While this may not seem a very intuitive picture 
at first, one of the advantages of describing the light 
field in terms of {\bf P} is that it directly refers to light field 
intensities as they are observed in experiment.

\subsection{Timescales and anisotropies}

The laser process in the split density model for quantum well VCSELs is
characterized by four timescales and three anisotropies. From slowest
to fastest, the timescales are defined by 
\begin{enumerate}
\item The rate of spontaneous emission into the laser mode $2w$, usually 
around $10^6 s^{-1}$ to $10^7 s^{-1}$, although this can be very much 
a function of cavity design \cite{Bab91,Hir96}.
\item The rate of spontaneous carrier decay $\gamma$, usually around 
$10^9 s^{-1}$ to $10^{10} s^{-1}$, which also depends strongly on the
cavity. In fact, the reduction of this decay into non-laser modes is
at the heart of a proposal to achieve thresholdless lasing by optimizing
the optical cavity \cite{Hir96}.
\item The rate of spin relaxation $\gamma_s$. This rate is an unknown
quantity, especially in the high carrier density regime crucial to the
laser structures under investigation. In this model, it appears only as
the sum with the spontaneous carrier decay which we define as
$\Gamma = \gamma_s + \gamma$.
Experiments and theory indicate, however, that the order
of magnitude will probably be similar to that of $\gamma$
\cite{Uen90}-\cite{Vin94}.
\item The rate of photon emission from the cavity $2\kappa$. 
\end{enumerate} 
The factor of 2 in the rate of spontaneous emission is a result of using 
photon number and intensity
variables instead of the field amplitudes to describe the light in the cavity. Also,
note that both $w$ and $\kappa$ are averages over anisotropic properties of
the VCSEL.

The three possible anisotropies in gain, loss and frequency are characterized
by both their magnitude and by their geometrical orientation. Similar to the
definition of the Stokes vector ${\bf P}$, it is possible to represent the 
anisotropies by vectors. The direction of this vector corresponds to the 
direction of {\bf P} for which the extremal gain/loss/frequency values are 
obtained. The length corresponds to the difference between the extremal 
values. The anisotropy vectors are defined as
\begin{enumerate}
\item gain anisotropy {\bf g}, such that the rate of spontaneous emission
  is given by $2w(1+{\bf Pg})$.
\item loss anisotropy {\bf l}, such that the rate of photon emission from the
cavity is given by $2\kappa(1+{\bf Pl})$.
\item frequency anisotropy ${\bf \Omega}$, such that the length of $\Omega$ is
equal to 1/2 the frequency difference between the modes of orthogonal 
polarization.
\end{enumerate}

\subsection{Rate equations}

We can now formulate the rate equations for any arbitrary set of anisotropies
using the parameters defined above. 
\begin{mathletters}
\begin{eqnarray}
\label{eq:rateeqns}
\frac{d}{dt}D &=& -w(1+{\bf Pg})Dn -\gamma(D-D_0)       -w(1+{\bf Pg})dnP_3\\
\frac{d}{dt}n &=&  w(1+{\bf Pg})Dn-2\kappa(1+{\bf Pl})n +w(1+{\bf Pg})dnP_3\\
\frac{d}{dt}d &=& -w(1+{\bf Pg})dn- \Gamma d  -          w(1+{\bf Pg})DnP_3\\
\frac{d}{dt}{\bf P} &=& [(w(1+{\bf Pg})(D{\bf g} + d{\bf e_3}) 
                        - 2\kappa(1+{\bf Pl}){\bf l})\times {\bf P}]\times{\bf P} \\
                    & &+({\bf \Omega} +w(1+{\bf Pg})\alpha d {\bf e_3})\times {\bf P} 
\end{eqnarray}
\end{mathletters}
${\bf e_3}$ indicates the unit vector in the direction of the 3rd component
of the stokes vector.

$D_0$ is the injection current in units of $\gamma$ and $\alpha$ is the 
linewidth enhancement factor, which describes a shift in frequency due 
to the electron-hole density in the quantum well. In the presence of 
two separate reservoirs of electron-hole pairs (split density model), where the carriers from
one reservoir interact only with one type of circular polarization, the frequency shift causes an
effective birefringence \cite{Mar96}.  

To understand the terms in the equation, it is important to keep
in mind that 
each time $\kappa$ or $w$ appear in the equations, they 
are modified by the anisotropy factors of $(1+{\bf Pl})$ and
$(1+{\bf Pg})$, respectively. 
In the model this is the only influence of the anisotropies on the dynamics of
$D$ and $n$ if $d=0$. 
If $d \neq 0$, the last terms in (2a) and (2b) increase or decrease the rate of
stimulated emission, depending on whether the light field in the cavity 
interacts more strongly or more weakly with the available carriers.
The last term in (2c) describes the hole burning effect of $P_3 \neq 0$. 

The dynamics of the Stokes parameters ${\bf P}$ is directly determined by the anisotropies. 
The frequency anisotropy induces a rotation around the axis defined
by ${\bf \Omega}$. The gain-loss anisotropy draws ${\bf P}$ towards
one of the poles along the axis defined by ${\bf g}$ and ${\bf l}$,
at a rate proportional to the component of ${\bf g}$ or ${\bf l}$ orthogonal
to ${\bf P}$. This is illustrated in Fig.~\ref{stokes}.

The effect of the split density model on the polarization dynamics 
can be understood in terms of gain and frequency anisotropies introduced by
the density difference $d$:
\begin{eqnarray}
{\bf g}_{eff}      & = & {\bf g}+\frac{d}{D}{\bf e_3}\\
{\bf \Omega}_{eff} & = & {\bf \Omega}+w(1+{\bf Pg})\alpha d{\bf e_3}.
\end{eqnarray}
In passing we note that another consequence  of the split density model is the fact that the factor
of 2 in the rate of spontaneous emission is absent.
This is due to the two parallel laser processes having an induced 
emission rate of $2w(D/2)(n/2)$ for $d=0$ and $P_3=0$. The sum of the two 
emission rates is therefore wDn, as given above.


\section{Langevin Equations for cw Operation at a stable linear Polarization}

\label{sec:Langevin}

\subsection{Anisotropies along the $[110]$ and $[1\bar{1}0]$ crystal axes}

Although a large number of nonlinear effects and dynamical properties can
be described by chosing special combinations of anisotropies
\cite{Tra96}
, we will concentrate on the simple case of cw operation at 
a stable linear polarization. For many VCSELs, this seems to be the 
natural state of affairs, even when no artificial anisotropies were 
created during the growth of the device. It has been found that
a large number of VCSELs emit light polarized along the $[110]$ or
$[1\bar{1}0]$ crystal axes
\cite{Cha91} - \cite{Jan96}
. This must obviously be the result of unintentionally introduced 
anisotropies. Possible reasons for this are the slight tilt of the 
growth axis often used in MOCVD
\cite{Cho95}
or the tendency of strain induced changes in the optical properties 
to produce anisotropies along these axes
\cite{Jan96}
. The latter effect can be visualized quite nicely. If one looks at 
a $[001]$ surface of a semiconductor lattice, the projections of the
bonds appear along the $[110]$ and the $[1\bar{1}0]$ directions. Light
polarized along one of these directions will mainly interact with the
electrons in the bonds along this direction. Consequently, the largest part
of the optical anisotropy induced by stress is caused by the difference 
in compression between the $[110]$ and the $[1\bar{1}0]$ bond directions.

Given this preference for two orthogonal linear polarizations observed in
many VCSELs, we choose ${\bf g} = g{\bf e_1}$,  ${\bf l} = l{\bf e_1}$
and  ${\bf \Omega} = \Omega{\bf e_1}$. The rate equations are now
\begin{mathletters}
\begin{eqnarray}
\label{eq:rateeqn2}
\frac{d}{dt}D &=& -w(1+ P_1g)Dn-\gamma(D-D_0)  -w(1+ P_1g)dnP_3\\
\frac{d}{dt}n &=& w(1+P_1g)Dn-2\kappa(1+P_1l)n +w(1+ P_1g)dnP_3\\
\frac{d}{dt}d &=& -w(1+ P_1g)dn-\Gamma d  - w(1+ P_1g)DnP_3\\[2ex]
\frac{d}{dt} P_1 &=& -(w(1+ P_1g)Dg-2\kappa (1+ P_1 l) l)(P_1^2-1) 
                     -w(1+ P_1g)d(P_3 P_1 - \alpha P_2)\\
\frac{d}{dt} P_2 &=& -(w(1+ P_1g)Dg-2\kappa (1+ P_1 l) l)P_1 P_2 
                     -w(1+ P_1g)d(P_3 P_2 + \alpha P_1) - \Omega P_3\\
\frac{d}{dt} P_3 &=& -(w(1+ P_1g)Dg-2\kappa (1+ P_1 l) l)P_1 P_3
                     -w(1+ P_1g)d(P_3^2-1)+ \Omega P_2.
\end{eqnarray}
\end{mathletters}

\subsection{Linearization around the stationary solution}

The stationary solution of these equations is given by $P_1=1$,
$P_2 = P_3 = d = 0$ and $D=2\kappa(1+l)/w(1+g)$. The photon number
in the cavity $n_s$ is a linear function of $D_0$, specifically
\begin{equation}
n_s=\frac{\gamma}{2\kappa (1+l)} D_0 - \frac{\gamma}{w(1+g)}.
\end{equation}
Close to this stationary point the laser relaxation dynamics can be
linearized. Deviations from the stationary point are described by
five coupled dynamical variables.  
In the case under consideration 
the dynamics can be seperated into two mutually decoupled sub-systems. 
One sub-system describes the dynamics of the total photon number 
fluctuations $\delta n=n-n_s$ coupled to the total density fluctuations  
$\delta D = D - 2\kappa(1+l)/w(1+g)$. 
The other sub-system describes the coupling of the density difference 
$d$ to the polarization parameters $P_2$ and $P_3$. 

To obtain the linearized Langevin equation, we further add the noise term
{\bf f}(t) to the dynamics. This time dependent five dimensional vector 
incorporates all external noise and signals influencing the laser. In the
case under consideration, this will be the vacuum fluctuations of the light
field entering the cavity. A discussion of the statistical properties of 
this noise term will be given in the next subsection.

At this point, it is convenient to introduce five new parameters for a 
more compact formulation:
\begin{enumerate}
\item The injection current in units of the threshold current is defined 
as x, which means that the stationary photon number $n_s$ is replaced by
\begin{equation}
x=\frac{w(1+g)}{\gamma}n_s +1.
\end{equation}
\item The gain-loss anisotropy is combined into a single variable $\rho$ 
equal to the difference of the two, scaled with the ratio of cavity loss rate
and spontaneous carrier decay. This dimensionless quantity should be of the 
order
of unity to be effective, corresponding to a relative anisotropy of 
about 0.1 percent for typical timescales:
\begin{equation}
\rho = \frac{2\kappa (1+l)}{\gamma}(g-l).
\end{equation}
\item The frequency anisotropy is scaled in terms
of the spontaneous carrier decay rate divided by $\alpha$, as it only 
stabilizes
the polarization in conjunction with $\alpha$ \cite{Mar96}:
\begin{equation}
\theta = \frac{\alpha \Omega}{\gamma}.
\end{equation}
\item The ratio of spin relaxation rate and spontaneous carrier decay rate
is written as $r$, such that
\begin{equation}
r=\frac{\Gamma}{\gamma}-1.
\end{equation}
\item The relaxation oscillation frequency $\nu$ is defined as a function of 
the injection current $x$:
\begin{equation}
\nu = \sqrt{ 2\kappa (1+l) \gamma (x-1)}
\end{equation}

\end{enumerate}

Using these parameters, the Langevin equation for the quantum well VCSEL is
\begin{equation}
\frac{d}{dt}\left(
\begin{array}{c}
\delta D\\ \delta n \\d\\P_2\\P_3
\end{array}\right) = \left(\! 
\begin{array}{ccccc}
-\gamma x & -\frac{\nu^2}{\gamma(x-1)} & 0 & 0 & 0 \\
\gamma (x-1) & 0 & 0 & 0 & 0 \\
0 & 0 & -\gamma (x+r) & 0 & \frac{\nu^2}{w(1+g)}\\
0 & 0 & -w(1+g)\alpha & -\gamma\rho & -\frac{\gamma\theta}{\alpha} \\
0 & 0 & -w(1+g) & \frac{\gamma\theta}{\alpha} & -\gamma\rho
\end{array}\right)
\left(
\begin{array}{c}
\delta D\\ \delta n \\d\\P_2\\P_3
\end{array}\right) + {\bf f}(t)
\\[2ex]
\end{equation}
All processes described by this equation occur either at a rate of $\gamma$
or at a rate of $\nu$. Since it is realistic to assume that $\gamma \ll \kappa$, 
the spontaneous carrier decay rate $\gamma$ will usually be significantly smaller 
than the relaxation oscillation frequency $\nu = \sqrt{ 2\kappa (1+l) \gamma (x-1)}$.
Note that this condition breaks down very close to threshold ($x=1$). 
However, as close to threshold the relative polarization fluctuations approach infinity
there is no polarization stability for $x$ very close to one.  

\subsection{Quantum noise}

Although the formulation given above can be applied also to problems of
externally injected fields and similar linear response problems, we will
now define the noise term ${\bf f}(t)$ as the electro-magnetic field noise
entering the cavity from the vacuum. 

The definition of a noise term for the light field in a cavity is a 
standard procedure in quantum optics. In the case of two modes of 
orthogonal polarization, the extension is 
straightforward. The two time correlation functions which are non-zero 
are
\begin{mathletters}
\begin{eqnarray}
\label{eq:non0corfunc}
\langle f_{\delta n}(t)f_{\delta n}(t+\tau)\rangle 
&=& 4\kappa(1+l)n_s\delta(\tau) \nonumber\\                                      
&=& \frac{4\kappa(1+l) \gamma (x - 1)}{w(1+g)}\delta(\tau)\\
\langle f_{p_2}(t)f_{p_2}(t+\tau)\rangle 
&=& \frac{4\kappa(1+l)}{n_s}\delta(\tau)\nonumber\\
&=& \frac{4\kappa(1+l)w(1+g)}{\gamma (x - 1)}\delta(\tau)\\
\langle f_{p_3}(t)f_{p_3}(t+\tau)\rangle 
&=& \frac{4\kappa(1+l)w(1+g)}{\gamma (x - 1)}\delta(\tau).
\end{eqnarray}
\end{mathletters}
To understand the derivation of these terms, one may either think in terms of 
a classical field, considering the vacuum modes and the dipole densities
associated with the carrier density pools as field modes with random 
quantum fluctuations acting as external forces on the field in the cavity, 
or one may
apply particle picture reasoning: since only whole photons are emitted
into and out of the cavity, there
is a stochastic process involved which gives rise to shot noise. It is one of
the fascinating properties of quantum mechanics, that the same noise terms 
result from two seemingly different pictures.

Another noise source for the laser process is the noise in the carrier 
injection. However, this noise is much weaker than the light field noise
if $\gamma x \ll \kappa$, which is a natural assumption for real devices. 
The physical reason for the relative smallness of shot noise from injection 
is that, if the photon emission rate from the cavity is much faster 
than the total carrier decay rate, then the number of
carriers present in the active region is far greater than the number of photons
in the cavity. Consequently, the relative statistical fluctuations 
are much smaller in the carrier subsystem than in the light field. 

Note that this approximation is only valid for short timescales. On 
timescales much longer than the $1/\gamma$, energy conservation
requires the fluctuations described here to cancel. This means that the
low frequency noise of the power spectrum is very sensitive to
the weak fluctuations in the carrier injection \cite{Yam86}.


\section{Solution of the Langevin Equation}

\label{sec:solution}

\subsection{The Greensfunction solution near the stationary point}

In accordance with the fluctuation-dissipation theorem, the noise in the
light emitted from the laser is approximately given by a linear response
to the quantum noise entering the laser cavity. The fluctuations can 
therefore be calculated from the linear response of the VCSEL. The 
five dimensional Greensfunction can be obtained by determining the 
Eigenvalues and Eigenvectors of the non-symmetric five by five matrix
describing the linearized dynamics. Since the Eigenvectors are non-orthogonal,
both left and right Eigenvectors need to be determined.

The problem separates into one two dimensional problem and one three
dimensional problem. Therefore, an exact analytical solution is possible. 
However, to
understand the physical significance and to single out the experimentally 
relevant case, 
it is useful to apply the realistic assumption that, since
 $\gamma x \ll \kappa $, $\gamma \ll \nu$, except for x very close to 1. 
Also, the anisotropies $\rho$ and $\theta$ should not be much greater than 1.
Using these assumptions, the Eigenvalues $\lambda_i$ and Eigenvectors 
${\bf a}_i$ and ${\bf b}_i$ are
\\[2ex]
\begin{math}
\lambda_{1/2} = -\frac{1}{2}\gamma x \pm i\nu 
\end{math} 
\begin{equation}
{\bf a}_{1/2}=\frac{1}{\sqrt{2}}
\left(
\begin{array}{ccccc}
\mp i \frac{\gamma (x-1)}{\nu} & 1 & 0 & 0 & 0 
\end{array}\right)\hspace{2em}
{\bf b}_{1/2}=\frac{1}{\sqrt{2}}
\left(
\begin{array}{c}
\pm i \frac{\nu}{\gamma (x-1)} \\1\\0\\0\\0 
\end{array}\right)
\end{equation}
\\[2ex]
\begin{math}
\lambda_3 = -\gamma (\rho+\theta)  
\end{math}
\begin{equation}
{\bf a}_{3}=
\left(
\begin{array}{ccccc}
0&0&0&1&-\alpha 
\end{array}\right)\hspace{2em}
{\bf b}_{3}=
\left(
\begin{array}{c}
0\\0\\0\\1\\0 
\end{array}\right)
\end{equation}
\\[2ex]
\begin{math}
\lambda_{4/5}=-\frac{1}{2}\gamma (x+r + \rho - \theta) 
\pm i \nu   
\end{math}
\begin{equation}
{\bf a}_{4/5}=\frac{1}{\sqrt{2}}
\left(
\begin{array}{ccccc}
0&0&\mp i \frac{w(1+g)}{\nu}&0&1 
\end{array}\right)\hspace{1em}
{\bf b}_{4/5}=\frac{1}{\sqrt{2}}\left(
\begin{array}{c}
0\\0\\ \pm i \frac{\nu}{w(1+g)} \\ \alpha \\ 1 
\end{array}\right).
\end{equation}
\\[2ex]
  
Using this set of vectors, any external perturbation of field variables
or carrier densities can be decomposed into the Eigenvectors with the
corresponding exponentially decaying and oscillating Greensfunction.
The time integral over the perturbation then gives the linear response of 
the laser dynamics. This procedure can now be applied to quantum noise.

\subsection{Application of the Greensfunction to quantum noise}

The left Eigenvectors ${\bf a}_i$ are used to decompose the noise into contributions 
associated with the corresponding Eigenvalues. If the noise terms are represented in the
form of a five by five diffusion matrix ${\bf N}$ \cite{Walls95book}, the decomposition  can be
accomplished by calculating the matrix elements $N_{i,j}={\bf a}_i {\bf N} {\bf a}^{\dagger}_j$. 
The fluctuation matrix ${\bf F}(\tau)$ can then be expressed as a sum over the  dyadic products of
the right Eigenvectors ${\bf b}_i$. 
\begin{equation}
{\bf F}(\tau) = \sum_{i,j} \frac{N_{i,j}}{-\lambda_i - \lambda^{\ast}_j} e^{\lambda^{\ast}_j \tau} 
{\bf b}_i \otimes {\bf b}^{\dagger}_j
\end{equation}
As the Eigenvectors are not orthogonal, there are contributions for $i \neq j$ with complex 
$- \lambda_i - \lambda^{\ast}_j$. Since we apply the approximation that $\gamma \ll \nu$, however,
the absolute value of these terms is much smaller than those with $i = j$ and we can obtain the
light field fluctuations by summing over the five terms with $i = j$ only. This effectively means
that the Eigenvectors fluctuate independently. The fluctuation matrix is then given by a sum over
projection operators which must be symmetric.

\subsection{Light field fluctuations}

The light field fluctuations of polarized light have three degrees of
freedom, given by the total intensity n, the ellipticity $P_3$ and the
direction of linear polarization $P_2$. Including possible correlations,
these fluctuations are described by a symmetric three by three matrix.
However, since the dynamics of the total total intensity is decoupled 
from the polarization dynamics, there is no 
correlation between the total intensity and the polarization. Therefore,
four fluctuation terms are sufficient to completely describe the fluctuations
in the laser light during cw operation: 
\begin{eqnarray}
\langle \delta n(t) \delta n(t+\tau)\rangle &=&
\frac{2\kappa \left(1+l\right)}{w\left(1+g\right)}\frac{x-1}{x}
e^{-\gamma x \tau /2} cos(\nu\tau)\\
\frac{\langle \delta n(t) \delta n(t+\tau)\rangle}{n_s^2} &=& 
\frac{A}{x(x-1)}
e^{-\gamma x \tau /2} cos(\nu\tau)\\
\langle P_3(t)P_3(t+\tau)\rangle &=&
\frac{A}{(x-1)(x+r+\rho-\theta)}
e^{-\gamma (x+r+\rho-\theta) \tau /2} cos(\nu\tau)\\
\langle P_3(t)P_2(t+\tau)\rangle &=& \alpha \langle P_3(t)P_3(t+\tau)\rangle \nonumber \\
&=&\frac{\alpha A}{(x-1)(x+r+\rho-\theta)}
e^{-\gamma (x+r+\rho-\theta) \tau /2} cos(\nu\tau)\\
\langle P_2(t)P_2(t+\tau)\rangle 
&=& \frac{\alpha ^2 A}{(x-1)(x+r+\rho-\theta)}
    e^{-\gamma (x+r+\rho-\theta) \tau /2} cos(\nu\tau)\nonumber\\
& & + \frac{A \left(1+\alpha^2\right)}{\left(x-1\right)\left(\rho+\theta\right)} 
    e^{-\gamma (\rho+\theta) \tau} 
\end{eqnarray}
The factor $A$ is a measure of the overall magnitude of noise. 
\begin{equation}
A=\frac{2\kappa \left(1+l\right) w\left(1+g\right)}{\gamma^2}
\end{equation}
Fig.~\ref{distrib} shows the noise distribution at $\tau = 0$ for a realistic choice of parameters.
Fig.~\ref{flucts} shows the polarization fluctuations as a function of $\tau$.

There is a correlation between the two fluctuations, mediated by 
$\alpha$, which is a typical feature of the two density model. The 
$\alpha$ factor converts the density difference fluctuations $d$ into
frequency difference fluctuations between left and right circular 
polarization. This fluctuating birefringence causes the direction of 
polarization to fluctuate in phase with the ellipticity.
The fluctuations of the carrier density have not been given here, since 
they are difficult to observe experimentally. They would be strongly 
correlated with the oscillating field terms, being 90~degrees out of
phase with respect to the fluctuations of the field.

Because of this correlation effect, fluctuations in the direction of 
polarization $P_2$ are always much stronger than fluctuations in the 
ellipticity $P_3$ if $\alpha$ is greater than 1. 
Indeed, the values of 2 to 6 given for $\alpha$ in the literature suggest 
a difference of almost an order of magnitude.

Another typical feature of the polarization fluctuations is that they
approach infinity close to threshold. This is an indicator that the light
emitted very close to threshold is still lamp like. A non vanishing
amount of laser emission is necessary to overcome the noise effects and to
stabilize both the intensity and the polarization. 

Although three unknown parameters, r, $\rho$ and $\theta$, enter into the
model, the polarization noise terms calculated here are defined by only 
two parameters, namely $\rho + \theta$ and
$r+\rho-\theta$. To fully separate the effects of spin relaxation and of
anisotropy, one additional parameter is needed. This additional parameter
may be found by taking a closer look at the nearly degenerate oscillation 
frequencies in the intensity and in the polarization.

\subsection{Perturbation theory for the frequency of relaxation oscillations}

The relaxation oscillations appear equally in the total intensity and
in the polarization fluctuations. This is a direct result of the two 
density model: ellipticity fluctuations and the associated fluctuations in 
the polarization direction are a result of uncorrelated intensity 
fluctuations in the two circular polarizations. 

However, anisotropies couple the two subsystems and induce slight changes
in the frequency. By calculating the difference between the relaxation
oscillations in intensity and in polarization, further information on the 
anisotropies can be obtained.

The perturbative correction to the Eigenvalues may be obtained by 
calculating the matrix elements $M_{i,j}$ between the approximate 
Eigenvectors ${\bf a}_i$ and ${\bf b}_j$. The correction to the 
Eigenvalue $\lambda_i$ caused by a weak coupling
to the Eigenstates of $\lambda_j$ is given by $M_{i,j}M_{j,i}/(\lambda_i
-\lambda_j)$. Calculated to second order in $\gamma/\nu$, the Eigenvalue
corrections are
\begin{eqnarray}
\delta \lambda_1 &=& \frac{M_{1,2}M_{2,1}}{\lambda_1-\lambda_2} \\
&=& -i\frac{\gamma ^2 x^2}{8 \nu}\\
\delta \lambda_2 &=& - \delta \lambda_1
\end{eqnarray}
for the total photon number and
\begin{eqnarray}
\delta \lambda_3 &=& 0 \\
\delta \lambda_4 &=& \frac{M_{4,5}M_{5,4}}{\lambda_4-\lambda_5}
                     +\frac{M_{4,3}M_{3,4}}{\lambda_4-\lambda_3}\\
                 &=& -i\frac{\gamma ^2}{8 \nu}(x+r-\rho+\theta)^2 
                     + i\frac{\gamma ^2 \theta ^2 (\alpha ^2 +1)}{2 \nu}\\
\delta \lambda_5 &=& - \delta \lambda_4
\end{eqnarray}
for the polarization variables.
The difference in frequency between the intensity oscillations $\nu_n$ and
the polarization oscillations $\nu_P$ is 
\begin{equation}
\nu_n-\nu_P = \frac{\gamma ^2}{8 \nu}
           \left[ \left(x+r-\rho+\theta\right)^2
                 -x^2 -4\theta ^2 \frac{\alpha ^2 +1}{\alpha ^2}\right].
\end{equation}
Note that only a frequency anisotropy will make $\nu_P$ larger than $\nu_n$.
Therefore, if the polarization noise oscillates faster than the intensity 
noise, this is an indicator of a strong frequency anisotropy.


\section{Discussion of the Results and Experimental Possibilities}

\label{sec:discussion}

\subsection{Determination of timescales and anisotropies from the fluctuations
of the laser light}

The equations given above show the wealth of information that can be
obtained from measurements of the fluctuations in the laser light emitted
from a VCSEL during cw operation at a stable linear polarization.
 
Intensity noise is given by the relaxation rate $\gamma x/2$, the oscillation
frequency $\nu$ and the relative magnitude of $A/x(x-1)$. Experimental 
determination of these three quantities is equivalent to a measurement
of the three timescales $\kappa(1+l)$, $\gamma$ and w(1+g). 

The ellipticity noise $\langle P_3(t)P_3(t+\tau)\rangle$ differs
from intensity noise in the relaxation rate, which is 
$\gamma(x+r+\rho-\theta)/2$ instead of $\gamma x/2$, and in the relaxation 
frequency, although the latter is only slightly different. 
These timescales also show up in the correlation of ellipticity and 
polarization direction fluctuations, $\langle P_3(t)P_2(t+\tau)\rangle$, which is just
$\alpha$ times the ellipticity fluctuations. An observation of this 
correlation can provide not only strong evidence in support of
the split density model, but is also a direct measurement of $\alpha$.
Finally, the fluctuations of polarization direction, $\langle P_2(t)P_2(t+\tau)\rangle$,
include not only noise correlated to ellipticity fluctuations, but also
additional noise with a relaxation rate of $\gamma(\rho + \theta)$.
From the two relaxation rates and the difference in oscillation frequency,
the parameters r, $\rho$ and $\theta$ can be calculated, thereby allowing
an experimental determination of the spin relaxation rate, of the gain-loss
aniosotropy and of the frequency anisotropy. 

\subsection{A note on polarization stability}

The polarization fluctuations predicted by the split density model are 
largely fluctuations of the direction of linear polarization. 
Fluctuations of the ellipticity are smaller by at least a factor of 
$\alpha^2$. For very small anisotropies, the main contribution to the 
fluctuations will be from the weak relaxation rate of $\gamma(\rho+\theta)$,
implying an even greater discrepancy between fluctuations in ellipticity and
in polarization direction. 

To estimate the relative anisotropies necessary to stabilize polarization,
it is therefore most appropriate to require the condition $\rho + \theta \gg 
A(1+\alpha ^2)/(x-1)$ to be fulfilled. For reasonable estimates, $\rho$ and
$\theta$ should then be at least of order 1. This would necessitate a relative gain-loss 
anisotropy $g-l$ of at least $10^{-3}$ or a frequency anisotropy of at least 1~GHz. 

However, these values depend critically on the timescales of the laser
process, all of which enter into the magnitude of the fluctuations 
given by $A$. 

\subsection{Experimental possibilities}

To determine the timescales and the anisotropies of quantum well VCSELs
as described above,
time resolved measurements of the polarization fluctuations at a resolution
of at least picoseconds are required. Further, polarization filters for both
linear and circular polarization are needed. 

Ideally, $P_2$ and $P_3$ could be measured directly by separating the laser
beam using a birefringent material and measuring the intensity difference 
between the two parts. However, it is also possible to measure the 
fluctuations by inserting a filter and measuring the fluctuations in 
$n_s(1+P_{2/3})/2$. Since the fluctuations in n are not correlated with the
fluctuations in polarization, the resulting relative noise is just the 
avarage of total intensity noise and polarization noise.
E.g.~if one measures the fluctuations in the intensity of right circular
polarized light $I_+(t)$ one obtains
\begin{equation}
\frac{\langle I_+(t)I_+(t+\tau)\rangle}{\bar{I_+}^2}=
\frac{1}{2}
\left(
\frac{   \langle n(t)n(t+\tau)\rangle}{n_s^2} 
       + \langle P_3(t) P_3(t+\tau)\rangle
\right)
\end{equation}
A measurement of the correlation of ellipticity and polarization direction
fluctuations is very difficult, since it requires a splitting of the laser beam
without causing undesireable polarizations. Probably the best approach would
be to reflect only a small fraction of the beam at a right angle, using this 
light for the measurement of linear polarization direction and using the 
nearly unaltered beam for the measurement of ellipticity fluctuations.
Another possibility is to keep track of all the polarization properties of
the optical devices used, sorting out the contributions to the measured noise
afterwards. The Stokes parameters are quite convenient for this task, since
all polarization effects can be described by matrix multiplication.
  
In this manner, all timescales and anisotropies can be determined at a 
fixed injection current. If such an experiment works, it is also possible
to measure changes in these properties as injection current is increased.
A comparison with the temperature dependence at constant injection current
is also possible. This should reveal a lot of the material properties of
the semiconductor and help to clear up some of the open questions regarding
polarization switching.


\section{Conclusions and Outlook}

\label{sec:concl}

The analysis presented here shows how all the parameters of the two
density model for a VCSEL with stable linear polarization can be 
obtained from measurements of the polarization fluctuations.
Time resolved measurements of the polarization fluctuations in
VCSEL light can therefore unambiguously resolve open questions,
such as whether temperature effects cause polarization switching
or whether birefringence or gain-loss anisotropies are responsible
for polarization stability. 

This is extremely important, because
experimental results on polarization 
and intensity as a function of the injection current can only be 
interpreted when the correct mechanism of polarization stability 
has been identified. For example, the authors of \cite{Jan96} assume 
that only a frequency anisotropy contributes to polarization stability 
without considering the possibility that a
gain-loss anisotropy might have an effect as well.
Unverified assumptions are also the subject of the criticism
voiced in \cite{Tra96} about the explanations given in \cite{Cho94}.
Further experiments such as the ones proposed here are absolutely necessary to avoid
misleading interpretations.
 
The polarization fluctuations show features typical for the
split density model, which can be used as a test criterium for 
whether the model is valid in a given device or not.
Since the split density represents the effect of quantum well confinement 
on the polarization properties of VCSELs, this effectively tests the
quantum well structure in the active region.  
The quantity which depends strongly on the size of the quantum wells is
the spin relaxation rate. Experimental results on this rate are also of 
interest in connection with
calculations \cite{Uen90,Fer91} and luminesence experiments \cite{Vin94}
carried out to investigate spin flip scattering in quantum wells.

The rate equations presented here are formulated in a very general
way and may also be applied
to cases with more excotic anisotropies and timescales,
such as discussed in \cite{Mar96} and \cite{Tra96},
which show switching and/or include a magnetic field. 
In all these cases, the investigation of noise adds additional
predictions for experiment to the results and thereby increases
our understanding of the physics involved in the polarization
properties of VCSELs.


%
%

%

\begin{figure}
\caption{The arrows on the Poincare spheres illustrate the dynamical effect of 
anisotropies on the normalized Stokes vector ${\bf P}$. The sphere on the left
shows how the frequency anisotropy causes ${\bf P}$ to rotate around the axis
defined by the anisotropy. The sphere on the right shows how the gain-loss
anisotropy pulls ${\bf P}$ towards one of the poles defined by the vector
of anisotropy.}
\label{stokes} 
\end{figure}

\begin{figure}
\caption{Contour plot of the Gaussian Distribution corresponding to the polarization 
fluctuations at $\tau =0$ for $x=2, \alpha=2, r=2, \rho=2$ and $\theta=2$. 
This choice of parameters clearly shows the correlation between polarization 
direction and ellipticity. 
For $A = 0.01$, the fluctuations of 
$P_2$ correspond to deviation of approximately $\pm 5$ degrees in 
the direction of polarization.}
\label{distrib}
\end{figure}

\begin{figure}
\caption{Time dependence of the fluctuations 
$\langle P_i(t)P_j(t+\tau)\rangle$ 
for $x=2, \alpha=2, r=2, \rho=2$ and $\theta=2$. As explained in the text, the
fluctuation of polarization direction $(i,j)=(2,2)$ is largest, while the 
correlation
of polarization direction and ellipticity $(i,j)=(2,3)$ is exactly equal in 
magnitude to two times the
fluctuations in ellipticity $(i,j)=(3,3)$. Time is given in units of 
$\gamma^{-1}$, which is typically 100 ps to 1 ns. The variable A is typically
around 1/100.}
\label{flucts}
\end{figure}  
%

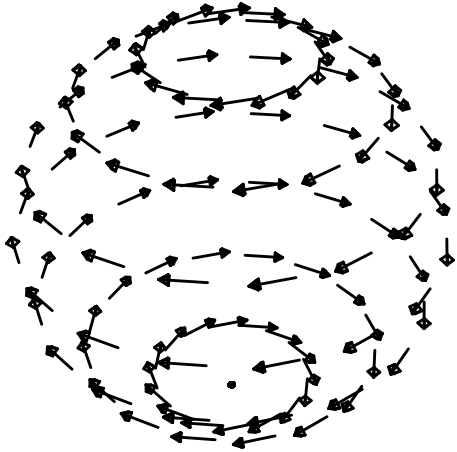
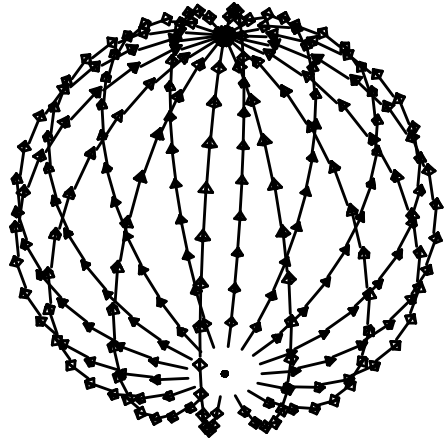
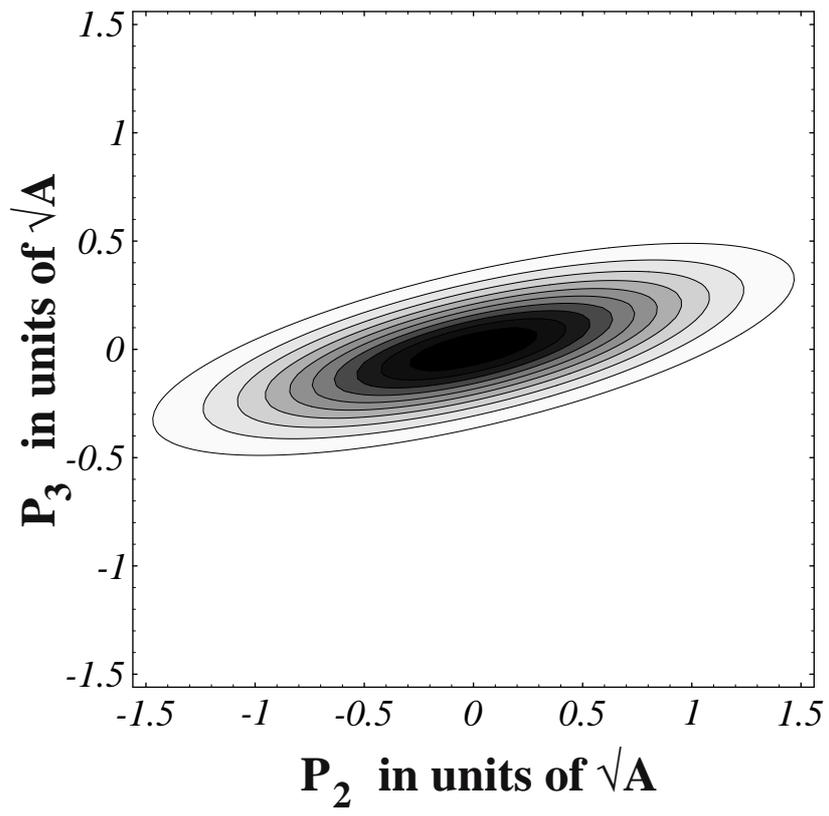
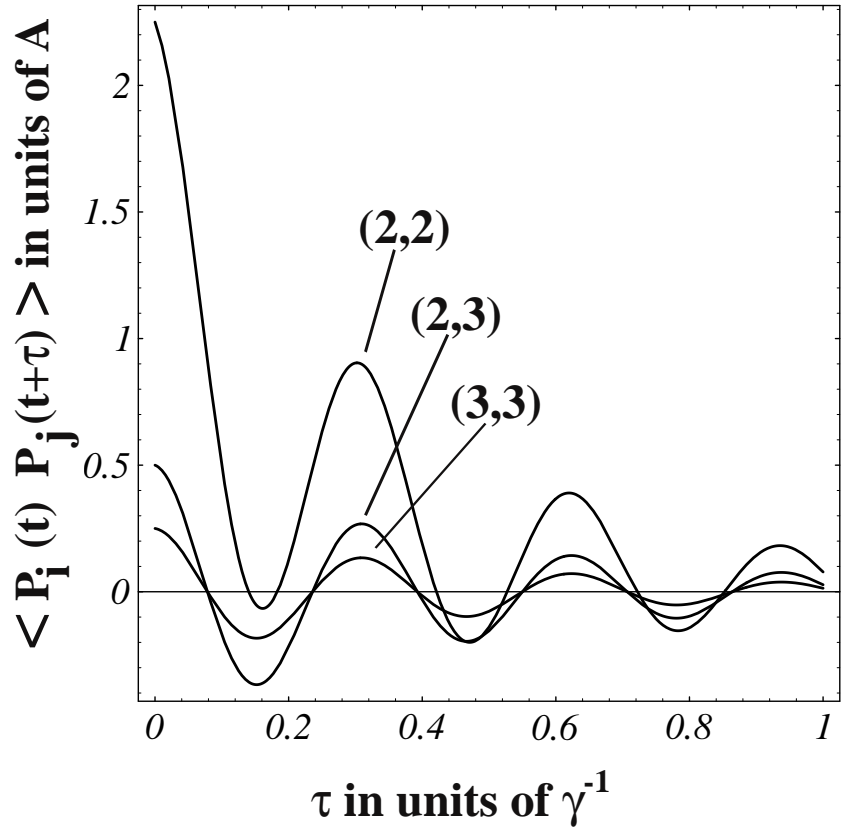
\end{document}